\documentclass[a4paper,11pt]{article}
\pdfoutput=1 

\usepackage{jcappub} 
\usepackage{xcolor}

\usepackage[T1]{fontenc} 
\newcommand{\sigmanu}{$\Sigma m_{\nu}$}
\newcommand{\summnu}{\Sigma m_{\nu}}

\newcommand{\lya}{Ly$\alpha$}
\newcommand{\lyaf}{Ly$\alpha$ forest}
\newcommand{\Mpc}{\ \text{Mpc}}
\newcommand{\hMpc}{\ h^{-1}\text{Mpc}}
\newcommand{\iMpc}{\ \text{Mpc}^{-1}}

\newcommand{\vx}{\mathbf{x}}
\newcommand{\vk}{\mathbf{k}}

\newcommand{\pvmhid}[1]{}

\title{Massive neutrinos and degeneracies in Lyman-alpha forest simulations}

\author[a]{Christian Pedersen,}
\author[a]{Andreu Font-Ribera,}
\author[b]{Thomas D. Kitching,}
\author[c]{Patrick McDonald,}
\author[d]{Simeon Bird,}
\author[e]{An\v{z}e Slosar,}
\author[f]{Keir K. Rogers,}
\author[a]{and Andrew Pontzen}

\affiliation[a]{Department of Physics and Astronomy, University College London,
London, UK}
\affiliation[b]{Mullard Space Science Laboratory, University College London,
Dorking, Surrey, UK}
\affiliation[c]{Lawrence Berkeley National Laboratory, One Cyclotron Road,
Berkeley, CA 94720, USA}
\affiliation[d]{Department of Physics \& Astronomy, University of California, 
Riverside, CA 92521, USA}
\affiliation[e]{Physics Department, Brookhaven National Laboratory, Upton, NY 11973, USA}
\affiliation[f]{Oskar Klein Centre for Cosmoparticle Physics, Stockholm University,
AlbaNova, Stockholm SE-106 91, Sweden}

\emailAdd{christian.pedersen.17@ucl.ac.uk}

\abstract{Using a suite of hydrodynamical simulations with cold dark matter, baryons, 
and neutrinos, we present a detailed study of the effect of massive neutrinos on the 1-D 
and 3-D flux power spectra of the Lyman-$\alpha$ (Ly$\alpha$) forest. The presence of 
massive neutrinos in cosmology induces a scale- and time-dependent suppression of 
structure formation that is strongest on small scales. Measuring this suppression is a 
key method for inferring neutrino masses from cosmological data, and is one of the main 
goals of ongoing and future surveys like eBOSS, DES, LSST, Euclid or DESI. The 
clustering in the \lyaf\ traces the quasi-linear power at late times and on small 
scales. In combination with observations of the cosmic microwave background, the forest 
therefore provides some of the tightest constraints on the sum of the neutrino masses. 
However there is a well-known degeneracy between \sigmanu\ and the amplitude of 
perturbations in the linear matter power spectrum. We study the corresponding degeneracy 
in the 1-D flux power spectrum of the \lyaf, and for the first time also study this 
degeneracy in the 3-D flux power spectrum. We show that the non-linear effects 
of
massive  neutrinos on the \lyaf, beyond the effect of linear power amplitude
suppression, are negligible, and this degeneracy persists in the 
\lyaf\ observables to a high precision. 
We discuss the implications of this degeneracy 
for choosing parametrisations of the \lyaf\ for cosmological analysis.}

\begin{document}
\maketitle

\flushbottom

\section{Introduction}
\label{sec:int}

The results of neutrino oscillation experiments show that at least two of the
neutrino mass eigenstates must have small but non-zero mass \cite{deSalas2018}.
Whilst these experiments are able to constrain the mass differences of the
eigenstates, they are far less sensitive to the absolute mass scale, or
to the sum of the mass eigenstates \sigmanu.
The extremely small cross section of neutrinos makes designing laboratory
experiments to measure their absolute mass scale challenging.
Attempts are continuing to be made through measuring the
$\beta$-decay spectrum of tritium \cite{BetaDecay,Katrin},
with the most recent results finding an upper limit
of $\Sigma m_\nu < 1.1$ eV (90\% C.L.) \cite{Katrin2019}.
The subtle effects of neutrino properties on cosmology have been studied for
decades \cite{BondEfstathiou1980,Julien06}, and the onset of precision data
sets in cosmology has opened up the possibility of measuring the neutrino mass
scale by detecting the effect on the growth of structure and expansion rate
of the Universe.

Constraints have been put on \sigmanu\ using several cosmological probes, such
as observations of the CMB \cite{Planck2018} or galaxy clustering and weak lensing
measurements from galaxy surveys \cite{DES2018}, and will continue to be a major
science goal in future surveys \cite{Hamann2012}.
Another cosmological probe that has emerged as an especially strong tool to
constrain neutrino mass is the clustering of the Lyman-$\alpha$ (\lya) forest:
a series of absorption features observed in the spectra of high redshift
($2<z<5$) quasars.
Some of the tightest constraints to date come from the combination of CMB and
\lya\ studies \cite{Seljak2006,PD2015}, with a current limit of only
$\summnu < 0.12$ eV (95\% C.L.).
With the upcoming Dark Energy Spectroscopic Instrument (DESI) survey
constraints are forecast to shrink to $\sigma_{\summnu}=0.041$eV when
utilising the full 3D power spectrum of the \lyaf\ and combined with data from
the CMB \cite{Font-Ribera2014}. Given the current lower limit from oscillation
experiments of $\Sigma m_\nu \geq 0.06$eV, these observations should begin 
to show evidence for neutrino mass.
When including information from the \lyaf\ bispectrum
the constraints could be further improved \cite{Mandelbaum2003}.

Three-dimensional correlations in the \lyaf\ have been measured at separations
of hundreds of Megaparsecs \cite{Busca2013,Slosar2013,Delubac2015,Bautista2017},
allowing a very precise determination of the expansion rate at $z \approx 2.4$
from baryon acoustic oscillations.
Most of the constraining power on total neutrino mass, however, comes from smaller
separations as massive neutrino free-streaming suppresses 
clustering on small scales at late times.
Current \lyaf\ constraints on neutrino mass are
restricted to the average correlation of one-dimensional Fourier modes along
quasar spectra, a summary statistic commonly known as the
1D flux power spectrum.

In order to extract cosmological information from the measured correlations
we need to be able to generate theoretical predictions for the flux power
spectrum as a function of cosmological model and as a function of several
nuisance parameters describing the uncertain thermal and ionization history
of the intergalactic medium (IGM).
In the absence of computing time constraints, one would run a large
hydrodynamical simulation for each of the $10^5-10^6$ likelihood evaluations
in a Monte Carlo Markov Chain.
However, these simulations are computationally expensive, typically requiring
$10^4-10^5$ core hours in high performance computing facilities, and past
\lyaf\ analyses were only able to simulate few tens of models.
Some studies used the simulations to calibrate a Taylor expansion of the
likelihood around a fiducial model \cite{Viel2006,Viel2010,Borde2014,Rossi2014,PD2015},
while others designed different interpolation frameworks to predict the flux
power spectrum for models that were not simulated \cite{McDonald2005}.
Recent works have used Gaussian processes to perform this interpolation,
which have the benefit of estimating an error on the prediction \cite{Bird2019}.
This allows for
the use of Bayesian optimisation to distribute the hydrodynamical simulations more
efficiently throughout parameter space, minimising the total number of 
simulations required and helping to ensure convergence to the true posterior distribution. \cite{Rogers2019}.

The analysis is further complicated by the existence of several parameter
degeneracies.
For instance, both the mean transmitted flux fraction (or mean flux) and the
amplitude of the linear power spectrum affect the overall amplitude of the
1D power spectrum of the \lyaf, and we are only able to break the degeneracy
because they affect its shape in a different way \cite{McDonald2005}.
These degeneracies are difficult to capture in an interpolating framework
or in a Taylor expansion, and numerical approximations might accidentally
break these degeneracies in the estimated likelihood.
Therefore it is important to choose a likelihood parameterization that minimizes the
parameter degeneracies.

In this paper we investigate the degeneracy between the sum of the neutrino
masses \sigmanu\ and the amplitude of the power spectrum.
This degeneracy has been studied in the context of the matter power
spectrum in real \cite{FVN2014b,Zennaro2019,Archidiacono2017} and redshift space \cite{FVN2018}.
However, as discussed in \cite{Viel2010}, the degeneracy is stronger in
studies of the 1D flux power spectrum of the \lyaf, since it is primarily sensitive to
the linear power on very small scales.

We revisit this degeneracy in the 1D flux power spectrum, with a
different simulation set up to previous studies that captures the degeneracy
more closely.
For the first time we also study the degeneracy in the 3D flux power spectrum.
In section \ref{sec:lin} we review the effect of massive neutrinos in
linear theory and discuss the parameter degeneracies.
In section \ref{sec:sim} we introduce a set of simulations to investigate
how well the degeneracy predicted by linear theory carries over into the
non-linear regime and into \lyaf\ observables. Here we also describe the
difference in our simulation set up when compared with previous studies
into the degeneracy in the \lyaf.
In section \ref{sec:res} we present our results, and we conclude in
section \ref{sec:con}.

\section{Linear theory}
\label{sec:lin}

In this section we review the effect of massive neutrinos on the linear
growth of structure (see \cite{Julien06} for a full review), focusing on 
effects on the linear power spectrum in the range of scales and redshifts
relevant in studies of the small-scale clustering of the \lyaf.
We denote the density parameters defined at $z=0$ for Cold Dark Matter (CDM),
baryons, neutrinos, and dark energy as
$\omega_\mathrm{c}$, $\omega_\mathrm{b}$, $\omega_\nu$ and $\omega_\Lambda$
respectively. For each component, these are related to the critical density
$\rho_c$ via the density fractions
$\Omega h^2=\omega$ where $\Omega=\rho/\rho_\mathrm{c}$.
The total non-relativistic
matter density at $z=0$ is
$\omega_{\mathrm{cb}\nu}=\omega_\mathrm{c}+\omega_\mathrm{b}+\omega_\nu$.

The first effect of a non-zero neutrino mass is a subtle change in the
expansion history of the Universe.
At early times, massive neutrinos are relativistic and indistinguishable from
massless ones, and their energy density decreases with the expansion of the
universe like radiation.
At late times, massive neutrinos become non-relativistic at redshift $z_\mathrm{nr}$
that is dependent on their mass, $m_\nu$:
\begin{equation}
1 + z_{\mathrm{nr}}= 189.4 ~ \left(\frac{m_\nu}{0.1 \mathrm{eV}}\right) ~.
\end{equation}
From this point their energy density evolves like non-relativistic matter, and the
total non-relativistic matter density is increased at the percent level:
\begin{equation}
 f_\nu \equiv \frac{\omega_\nu}{\omega_{\mathrm{cb}\nu}} = 0.023
    \left( \frac{\Sigma m_\nu}{0.3\ \mathrm{eV}} \right)
    \left( \frac{0.14}{ \omega_{\mathrm{cb}\nu}} \right) ~,
\end{equation}
where the neutrino energy density is given by
\begin{equation}
\omega_\nu = 0.00322 ~ \left( \frac{\Sigma m_\nu}{0.3 \mathrm{eV}} \right)~.
\end{equation}

\begin{figure}
  \centering
  \includegraphics[scale=0.5]{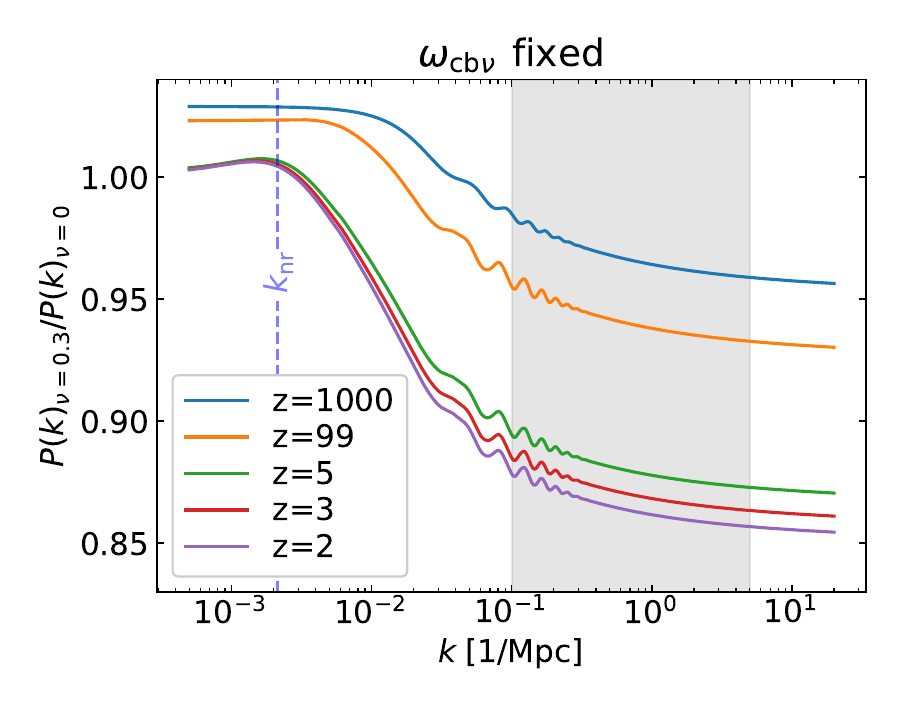}
  \includegraphics[scale=0.5]{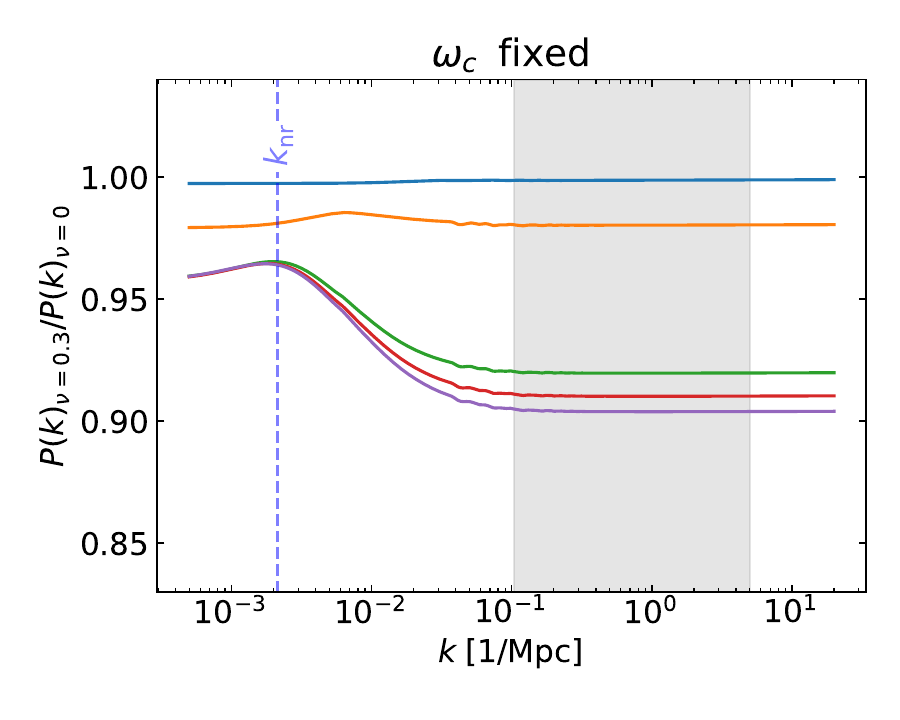}
  \caption{The effect of degenerate hierarchy $\Sigma m_\nu=0.3 \mathrm{eV}$ massive 
    neutrinos on the linear theory CDM + baryon power spectrum at several 
    different redshifts, in two different scenarios:
    (left) $\omega_c$ is reduced to conserve the value of $\omega_{\mathrm{cb}\nu}$ in both
    models;
    (right) $\omega_c$, $\omega_b$, and $h$ are kept fixed, with the
    increase in $\omega_{\mathrm{cb}\nu}$ when adding
    massive neutrinos compensated by 
    reducing the value of $\omega_\Lambda$ in the model.
    The suppression on small scales due to massive neutrino free streaming 
    is clear in both examples, however the small-scale suppression is only
    scale independent in the right panel.
    In blue dashed lines we show the neutrino free-streaming scale
    ,$k_{\rm nr}$, for a $0.1 \rm{eV}$ mass 
    eigenstate, and the shaded area approximately represents the length scales 
    probed by the one-dimensional flux power spectrum of
    the \lyaf.
}
  \label{fig:linear}
\end{figure}

After the non-relativistic transition neutrinos effectively behave like
hot dark matter in that they free-stream and do not cluster on small scales.
This length scale is set by the wavenumber which enters the horizon when neutrinos become non-relativistic:
\begin{equation}
\label{eq:freestream}
k_{\mathrm{nr}} ~ \simeq ~ 0.00213 \left( \frac{\omega_{\mathrm{cb}\nu}}{0.14} \right)^{1/2} \left( \frac{m_\nu}{0.10\mathrm{eV}} \right)^{1/2} \iMpc,
\end{equation}
This gives rise to the second effect of non-zero neutrino masses: the spatial
distribution of CDM and baryons is affected by the distribution of neutrinos,
as captured in the evolution of the linear power spectrum.
In figure \ref{fig:linear} we compare the linear power spectra
\footnote{Computed using the publicly available code {\texttt{CAMB}}\cite{CAMB}}
of CDM and baryons in a massless neutrino cosmology and a cosmology with
three $0.1{\rm eV}$ neutrinos. In this paper we assume that the \lyaf\
is sensitive to the combined CDM
and baryon power spectrum, and so we focus on this quantity rather than what is
traditionally referred to as the matter power spectrum and includes massive
neutrinos.

As can be seen in figure \ref{fig:linear}, the effect of massive neutrinos in
the linear power spectrum is both redshift and scale-dependent.
The scale dependence is set by the neutrino free-streaming scale
(eq. \ref{eq:freestream}), 
and we plot this as a vertical dashed line in figure \ref{fig:linear}.
When comparing massive and massless neutrino cosmologies, there are several
different choices one can make about which parameters to vary, and we show
two options in the panels.

In the left panel we change $\omega_c$ to keep constant the total density
of non-relativistic matter at low-redshift $\omega_{\mathrm{cb}\nu}$.
In the right panel we keep $\omega_c$ fixed and change the value of 
$\Omega_\Lambda = 1 - \Omega_{\mathrm{cb}\nu}$.
In both cases we keep fixed the baryon density parameter $\omega_b$
and the Hubble parameter $h$.
In the left panel the effect at $z=1000$ shown in blue is 
caused by the change in $\omega_c$, as this is well before
the neutrino relativistic transition ($z \gg z_{\mathrm{nr}} = 188$).
At low redshift, we see the characteristic suppression of power
below the neutrino free-streaming scale
($k \gg k_{\mathrm{nr}} = 0.0021 \iMpc$). We note that the
suppression has a mild scale- and 
redshift-dependence even on small scales ($k\approx 1 \iMpc$).

When we keep $\omega_c$ fixed (right panel), adding a non-zero mass to
neutrinos does not affect the physics of the early Universe 
($z \gg z_{\mathrm{nr}} = 188$), and the linear power at $z=1000$ 
is practically unaffected.
At much later times ($z \ll z_{\mathrm{nr}} = 188$), and on scales smaller
than the free-streaming scale, 
the effect of neutrinos is a scale-independent suppression of the linear 
power spectrum.
Studies of the small-scale clustering of the \lyaf\ are sensitive to the linear
power in the approximate range shown shaded in gray in figure \ref{fig:linear},
which is well below the neutrino free-streaming scale.
This suggests an almost perfect degeneracy in the linear power spectrum between
the effect of massive neutrinos and the amplitude of the primordial power
spectrum $A_s$ when considering only these length scales, redshift
ranges, and neutrino masses. For $\Sigma m_\nu>0.5 \mathrm{eV}$ the suppression
caused by neutrinos is no longer scale-independent at the lowest $k$ modes
probed by the \lyaf, and so further work would be required to generalise the
results presented in this paper for this neutrino mass range.
We note that this threshold is higher than the current upper bound of
$\Sigma m_\nu<0.24 \mathrm{eV}$ from CMB measurements alone \cite{Planck2018}.
In the remaining sections of this paper we use hydrodynamical simulations
to show that this degeneracy is also valid in the non-linear regime,
and is still very strong in the \lya\ flux power spectra.

\section{Simulations}
\label{sec:sim}

The simulations are run using the Tree-SPH code \texttt{MP-Gadget}
\footnote{\url{https://github.com/MP-Gadget/MP-Gadget}}, a modified version of
\textsc{Gadget-2} \cite{Gadget2} with the ability to simulate massive neutrinos
and with an improved performance in massively parallelized runs \cite{mp-gadget}.

Initial conditions for all our simulations were generated at $z=99$, using
separate, species specific, transfer functions for CDM and baryons computed using
{\texttt{CLASS}} \cite{CLASS}.
The CDM and baryon particles are both initialised on regular grids, offset to
prevent particles being initialised at the same position.
The initial conditions are generated using the Zel'dovich approximation at $z=99$
with {\tt MP-Gadget}'s inbuilt initial condition code, {\tt MP-GenIC}.
We do not use 2LPT because the terms have only been computed for a single
initial fluid.
The main results presented in this paper are from a set of simulations with box
size $L=133.85 \Mpc$ ($h\,L= 90 \Mpc$), and $1024^3$ CDM and baryon particles,
and we show in appendix \ref{app:box} that the conclusions of this paper do
not depend on box size or resolution. In order to reduce cosmic variance,
we use the `paired and fixed' simulations introduced in \cite{Angulo16,Anderson18}.
In this procedure, the initial amplitudes in each Fourier mode are fixed to the
ensemble average instead of being randomly drawn, and for each cosmological
model two simulations are run with the phases in each mode inverted.
Clustering statistics such as the power spectrum are then taken to be the average
of those calculated in each of the two simulations.
To include the effects of reionisation, we use a uniform UV background following the
model presented in ref.~\cite{Puchwein2019}, which has been tuned to approximately
match the observed IGM thermal history.

Including massive neutrinos in cosmological simulations presents its own
set of technical challenges \cite{Hannestad2012,Banerjee2016,Emberson2017,Bird2018,Dakin2019}.
Neutrinos are often included as another species
of particle in the simulation alongside the CDM and baryons 
\cite{Bird2012,FVN2014,Castorina2014}. However due to
the comparatively low clustering of neutrino particles, especially for the
lighter neutrino masses considered as the upper limit on $\Sigma m_\nu$ becomes
tighter, it is necessary to include a large number of neutrino particles in the
simulation in order to reduce the shot noise below the level of the physical neutrino 
clustering \cite{Wang2007}.
This is in turn more computationally intensive, adding $~50\%$ to the walltime of a 
given simulation when using the same number of neutrino particles as CDM,
as well as increasing memory and storage requirements.

An alternative approach was proposed in \cite{Brandbyge2009}, which we refer to as 
the {\it Fourier space} approach, where the neutrino clustering
is calculated using linear theory and then included in the gravitational potential
used to evolve the baryon and CDM particles. This technique was further improved to a
linear response in \cite{Haimoud2013}, where the linear evolution of the neutrino component is
determined using the full non-linear density field of the CDM and baryon particles
in the simulation. We use this implementation for the results presented in the
main text of this paper, and demonstrate in appendix \ref{app:nu} that our findings
do not depend on which approach is used.
Unlike the particle implementation, the storage and memory requirements for massive
neutrino simulations using the linear response approach
are very similar to massless neutrino simulations, with a $~5\%$ increase 
in walltime with respect to the massless case that is largely independent of
cosmological parameters.

The computation time of our simulations is significantly reduced by turning
regions with gas density of $\rho_b/\bar{\rho}_b > 10^3$ directly into stars
({\tt QuickLymanAlpha} option in Gadget simulations \cite{Viel2004}),
since these large overdensities
do not affect the \lyaf\ observables but are extremely computationally costly
to evolve.
The flux skewers are computed using \texttt{fake\_spectra}
\footnote{\url{https://github.com/sbird/fake\_spectra}}, which calculates the
optical depth along a given line of sight.
We compute the flux along a regular grid of $600^2$ lines of sight at
a cell resolution of $10 \mathrm{km/s}$,
providing high resolution both along and traverse to the line of sight.
To calculate the 1D flux power spectrum, we take the Fourier transform of
flux perturbations along each skewer, and then average each Fourier mode
across all skewers in a given simulation. For the 3D flux power spectrum we utilise
the fact that the skewers are calculated in an evenly spaced grid with the same
line of sight resolution, and take a Fourier 
transform of flux perturbations for the entire box.
The results are then averaged in bins of $k=|\vec{k}|$
and $\mu$, where $\mu$ is the cosine of the angle of each Fourier mode
with respect to the line of sight \cite{Rogers2017a,Rogers2017b}.

In this study we run simulations for three different cosmologies:
the {\it massive} simulation uses a cosmology with massive neutrinos
($\summnu=0.3$eV);
the {\it massless} simulation uses a similar cosmology but with massless
neutrinos, resulting in a slightly lower $\omega_{\mathrm{cb}\nu}$ 
at low redshift;
the {\it rescaled} simulation uses the same cosmology as the
{\it massless} simulation, but with a slightly lower amplitude of primordial
fluctuations $A_s$.
The value of $A_s$ in the {\it rescaled} simulation is chosen in order to
match the amplitude of the linear power spectrum of CDM + baryons in the 
{\it massive} simulation at a
central redshift $z=3$, and at a wavenumber\footnote{The exact scale where we 
match the linear power is not important,
see the right panel of figure \ref{fig:linear}.} $k=0.7 \iMpc$.
Parameters for these simulations are shown in Table \ref{tab:sims}.
The cosmology in the {\it massive} simulation has three neutrinos of
degenerate hierarchy and $\Sigma m_\nu=0.3\mathrm{eV}$, which is slightly
larger than the upper constraint provided by Planck alone \cite{Planck2018}.
We choose this extreme case as the idea we present will only become more
reliable with lower values of \sigmanu.

The first detailed study into the degeneracy between the amplitude of the
small scale linear power spectrum and $\Sigma m_\nu$ in the context of the
\lyaf\ was presented in \cite{Viel2010}.
In our simulation set up we build upon this work, and intend
to capture the degeneracy more completely with the following changes.
Firstly, when adding massive neutrinos
they kept the total matter content $\Omega_{\mathrm{cb}\nu}$ fixed,
which results in a slight
scale dependence in the suppression even on small scales as seen 
in the left panel of figure \ref{fig:linear} \cite{Viel2010,FVN2014}.
The effect of this is that the linear
theory power spectrum will not match on all length scales relevant for the \lyaf. 
Secondly, in order to mimic the effect of massive neutrinos, they match $\sigma_8$ at 
$z=7$, where $\sigma_8$ is calculated in each case from the total matter power
spectrum including massive neutrinos. The \lyaf\ is primarily sensitive to the clustering
of the baryons and CDM rather than the clustering of the neutrinos themselves,
and so we instead opt to match the CDM + baryon power spectrum at $z=3$, neglecting the
contribution from massive neutrinos.
The suppression also has a slight redshift dependence, so fixing $\sigma_8$ at
$z=7$ would still result in imperfect agreement between the linear theory
power spectra at lower redshifts where the \lyaf\ is observed.

\begin{table}
	\centering
  \begin{tabular}{ | l | l | l | l |}
  \hline
     & {\it massive} & {\it massless} & {\it rescaled} \\ \hline
     $\Sigma m_{\nu}$ (eV) & 0.3 & 0.0 & 0.0 \\
     $\Omega_{\mathrm{cb}\nu}$ & 0.3192 & 0.3121 & 0.3121 \\ 
     $A_s$ & 2.142e-9 & 2.142e-9 & 1.952e-9 \\ \hline
     $\Omega_c$ & \multicolumn{3}{c|}{0.2628} \\ 
     $\Omega_b$ & \multicolumn{3}{c|}{0.0493} \\ 
     $n_s$ & \multicolumn{3}{c|}{0.9667} \\
     $h$ & \multicolumn{3}{c|}{0.6724} \\ \hline
  \end{tabular}
  \caption{Simulations from which we obtain key results.
    We run three types of simulation: a simulation with $\Sigma m_\nu=0.3\mathrm{eV}$, 
    a massless neutrino simulation, and a massless neutrino
    simulation where $A_s$ has been rescaled in order to replicate the effect
    of massive neutrinos, referred to as a {\it rescaled} simulation.
    All cosmological parameters are kept constant except $A_s$,
    $\Sigma m_{\nu}$, and consequently $\Omega_{\mathrm{cb}\nu}$.}
  \label{tab:sims}
\end{table}

In the next section we compare the non-linear power spectra and \lyaf\
observables for the simulations described in Table \ref{tab:sims}, and study
to what degree the effects of massive neutrinos on these quantities can be
replicated.

\section{Results}
\label{sec:res}

In this section we present the results of our simulations. 
We examine how well the degeneracy in linear theory continues
into the non-linear regime by looking at the matter power spectra in the
three simulations described in Table \ref{tab:sims}.
We then study the effect of massive neutrinos on the \lyaf\
observables, the 1D and 3D flux power spectrum, as these are the key statistics
that are ultimately used to constrain cosmology.
As well as being dependent on the underlying non-linear power spectrum, the 
\lyaf\ is also affected by the thermal and ionisation states of the IGM, which we will discuss in closer detail in section
\ref{ss:lya}.

\subsection{Non-linear growth of structure}
\label{ss:non}

\begin{figure}
  \centering
  \includegraphics[scale=0.6]{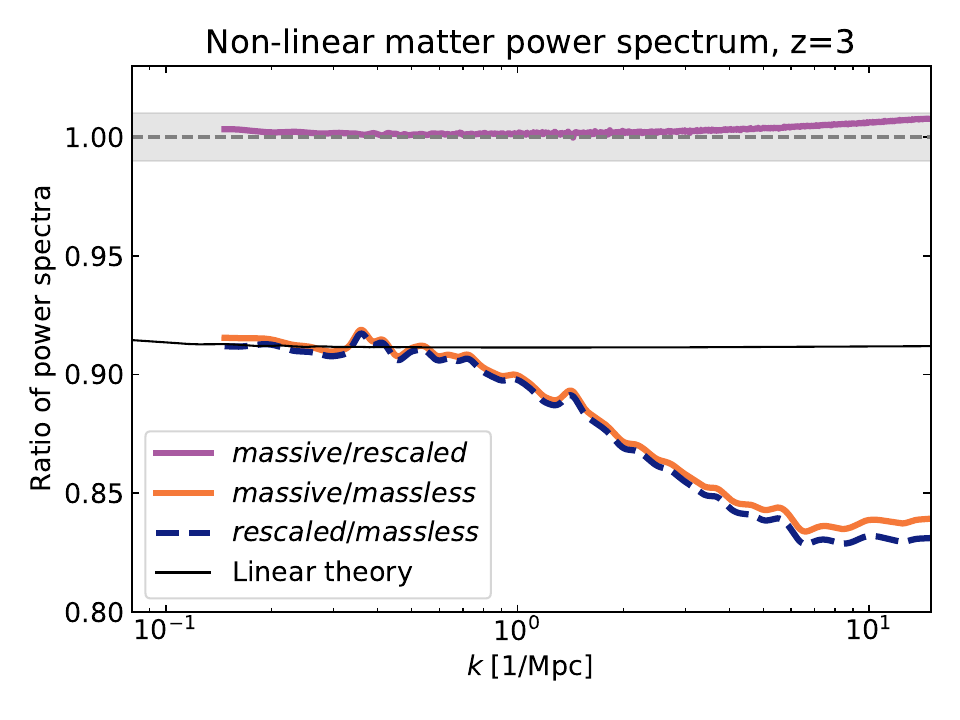}
  \caption{The ratio of the matter power spectra (CDM+baryons) in simulations
    with massive and massless neutrinos.
    The black line shows the linear theory result for the cosmologies in the
    {\it massive} and {\it massless} simulations described in Table \ref{tab:sims}
    ,which is the same as the red line in figure \ref{fig:linear} right panel.
    The solid orange line shows a comparison between the full non-linear matter
    power spectra in the {\it massive} and {\it massless} simulations.
    The solid purple line represents the comparison between the {\it massive}
    and {\it rescaled} cosmology, a massless neutrino cosmology with a lower
    clustering amplitude $A_s$ to match the small-scale linear power at $z=3$.
    In dashed blue, we compare the two massless neutrino cosmologies, showing
    that the `spoon' effect can be recreated without any massive neutrinos.
    The gray band shows the region of $1\%$ agreement.}
  \label{fig:matpow}
\end{figure}

First we look at the effect of massive neutrinos on the growth of structure in
the non-linear regime, and examine to what extent these effects can be
replicated simply by rescaling $A_s$.
In figure \ref{fig:matpow} we plot the ratio of the CDM + baryon matter power
spectra at $z=3$ in our three cosmologies:
The black line shows the linear theory result, also shown in the right panel
of figure \ref{fig:linear}.
The solid orange line compares the power in {\it massive} and {\it massless}
simulations, where $\Omega_c$ and $\Omega_b$ are kept constant,
and a $\summnu=0.3\mathrm{eV}$ neutrino mass has been added changing
$\Omega_{\mathrm{cb}\nu}$.
We see the characteristic `spoon' effect also seen in \cite{Viel2010,Bird2012}
and references therein, which is because the onset of non-linearities are delayed
as a result of the suppression of linear power.

The solid purple line shows the ratio of the power spectra in
massive and massless neutrino
cosmologies; however in this case we have rescaled the perturbation amplitude
of the massless cosmology to match the small-scale linear power at $z=3$
({\it rescaled} cosmology in Table \ref{tab:sims}).
The matter power spectrum at $z=3$ in these two simulations agree to within
$1\%$ on all length scales relevant for \lyaf\ analysis.
We note that the star formation in these simulations is also
extremely similar, with a difference of 0.05\% at $z=2$.
The blue dashed line shows the ratio of the power spectra in the two
massless cosmologies described in
Table \ref{tab:sims}, that differ only in the value of $A_s$.
In this case we see the same spoon effect as in the massive neutrino
comparison, highlighting the fact that this feature is a consequence of the
suppression of structure growth on linear scales and can be replicated
without massive neutrinos.

If the \lyaf\ power spectrum were measured at a single redshift, 
this would be sufficient: the solid purple line in figure \ref{fig:matpow}
shows that the effect of massive neutrinos can be reproduced to sub-percent
agreement by a rescaling of the amplitude of initial perturbations.
However, in the right panel of figure \ref{fig:linear} we see that the effect
of massive neutrinos on the small scale linear power has a redshift
dependence, varying by a couple of percent over the redshift range covered
by the \lyaf, $5 > z > 2$, meaning that rescaling $A_s$ will only match
the linear power perfectly at a single redshift.

\begin{figure}[t!]
  \centering
  \includegraphics[scale=0.6]{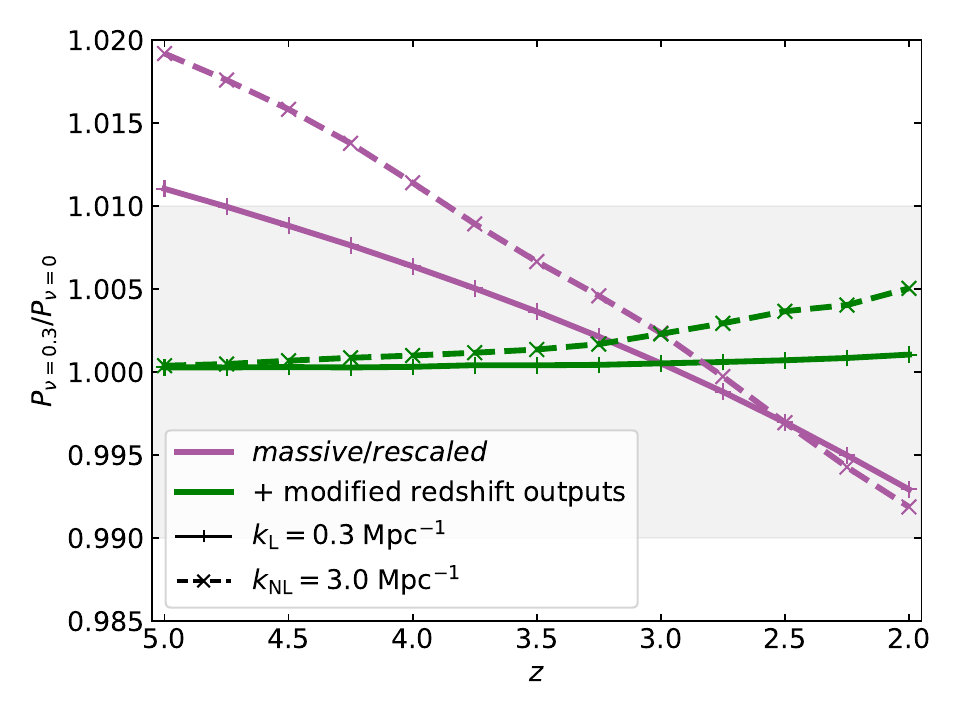}
  \caption{The ratios of the matter power spectra in the {\it massive}
  and {\it rescaled} simulations for a linear (solid lines) and non-linear
  (dashed) mode across the full redshift range $5>z>2$. In purple lines, the
  ratio has a small redshift dependence due to the fact that the growth
  in the two cosmologies is different. By design, the ratios in the linear
  modes are matched at $z=3$. In green lines, we have modified the redshifts
  at which snapshots are output to match the linear power in the two simulations.
  When the linear power is matched, the power in the non-linear mode agrees
  to within $0.5\%$.
}
  \label{fig:relabel}
\end{figure}

To investigate this effect we pick two modes, one linear ($k_{\rm L} = 0.3 \iMpc$)
and one non-linear ($k_{\rm NL} = 3 \iMpc$). In the purple lines of figure \ref{fig:relabel},
we plot the ratio of the power in these modes in the {\it massive} and {\it rescaled}
simulations, with the ratio for $k_{\rm L}$ shown in solid lines, and
$k_{\rm NL}$ shown in dashed lines. By construction the amplitude in the linear
mode is matched at $z=3$, but due to the different growth rates in the two
cosmologies, there are disagreements at the percent level at $z=5$ and $z=2$.
We note that this is a very small effect that is still well below the precision
of current measurements of the amplitude of the linear power from the \lyaf\, which
are on the order of $10\%$ \cite{McDonald2005,Chabanier2019}. This effect is also very small
when compared with the size of the suppression caused by massive neutrinos, which
is $9\%$ for our case of $\Sigma m_\nu=0.3$.

In the green lines we plot the same ratios, except where we have output the snapshots
in the {\it rescaled} simulation at a slightly different redshift in order to match
the amplitude of the linear power.
For instance, we compare the snapshot at $z=5$ for
the {\it massless} simulation with a snapshot of the {\it rescaled} simulation at $z=4.97$.
While the ratio of the power in the linear modes changes as a function of redshift
in the purple lines, it is constant in the green lines. When matching the amplitude
of the power in the linear mode, the power in the non-linear mode agrees to within
$0.5\%$ across the full redshift range, even when massive
neutrinos are not included in the simulation. This tells us again that the non-linear structure
is primarily sensitive to the amplitude in the linear power, and that any non-linear
effects caused by massive neutrinos themselves are negligible with respect to current
precision.

\subsection{Lyman-$\alpha$ forest clustering}
\label{ss:lya}

In this section we look at the statistics of the \lyaf, and examine the
degeneracy between massive neutrinos and the amplitude of primordial
fluctuations.
In particular, we look at correlations of the fluctuating transmitted flux
fraction, $ \delta_F (\vx) = F (\vx) / \bar{F} -1$, where $\bar{F}$ is the
mean transmitted flux fraction. 
In principle, one could measure the full 3D power spectrum,
\begin{equation}
  \left\langle \delta_F(\vk) \delta_F(\vk') \right\rangle 
    = (2 \pi)^3 \delta^D(\vk+\vk') P_{3D}(k,\mu)
\end{equation}
where $\delta_F(\vk)$ is the Fourier transform of $\delta_F(\vx)$ and $\mu$ is
the cosine of the angle between the Fourier mode and the line-of-sight, and
use it to constrain cosmology \cite{Font-Ribera2018}.
However, current \lya\ constraints on neutrino mass use exclusively the
1D power spectrum, 
\begin{equation}
  P_{\mathrm{1D}}(k_\parallel) = \int_{0}^{\infty} \frac{dk_\perp k_\perp}{2\pi} 
    P_{\mathrm{3D}}(k_\perp,k_\parallel) ~,
\label{eq:p1d}
\end{equation}
where $k_\parallel$ and $k_\perp$ are the components of the Fourier mode along
and perpendicular to the line of sight respecitvely,
$k_\parallel=k \mu$, and $k^2 = k_\parallel^2 + k_\perp^2$.

The 1D and 3D flux power spectra of the \lyaf\ are dependent
on both the matter density field and the state of the IGM, and there is also
some dependence on the velocity power spectrum through redshift space distortions.
Due to the different initial conditions, the IGM history in the {\it massive} and
{\it rescaled} simulations is different, which will propagate to differences in the
flux power spectra. However these differences do not mean that the degeneracy
shown in figures \ref{fig:matpow} and \ref{fig:relabel} is broken as the state of the
IGM is marginalised over in cosmological analysis of the \lyaf\ due to uncertainties
in the astrophysics of reionisation. These subtle changes in the state of the IGM
therefore cannot be interpreted as signatures of massive neutrinos.
Given imperfect parametrisations of the IGM and the highly non-linear relationship
between these parameters and the observable flux power spectra, it is impossible to
isolate all of these effects and match them by hand in the way that we have done with
the linear power.

A full marginalization is beyond the scope of this paper, but we match two
of the IGM parameters -- the mean flux and the temperature at mean density in the
simulations -- to remove some effects of neutrinos that are degenerate with
nuisance parameters. The temperature of the gas in the IGM is well approximated by
a power-law distribution of the form $T(\delta_b) = T_0 (1 + \delta_b)^{\gamma-1}$
where $\delta_b$ is the baryon overdensity, and $T_0$ is the temperature
at mean density \cite{Hui1997}. Given that the power-law approximation
only holds at low densities, we perform the fit in the range
$-2 < \rm{log}_{10}(\delta_b)<0.5$. This is appropriate given that these
overdensities are also the regions which give rise to the \lyaf\ \cite{Lukic2014}.
We match the mean flux by re-scaling the optical depth in the skewers by a
constant in post-processing. 
We also match $T_0$ in the simulations. For the heavily ionised gas in
photoionisation equilibrium which gives rise to the \lyaf\, the gas temperature is 
proportional to the internal energy. This allows us to rescale the temperature at mean 
density in the simulations by a constant rescaling of the internal energies of the 
particles, as long as the rescalings are small. In both cases the rescalings were 
small, on the order of 2\%.

\begin{figure}[t!]
  \centering
  \includegraphics[scale=0.47]{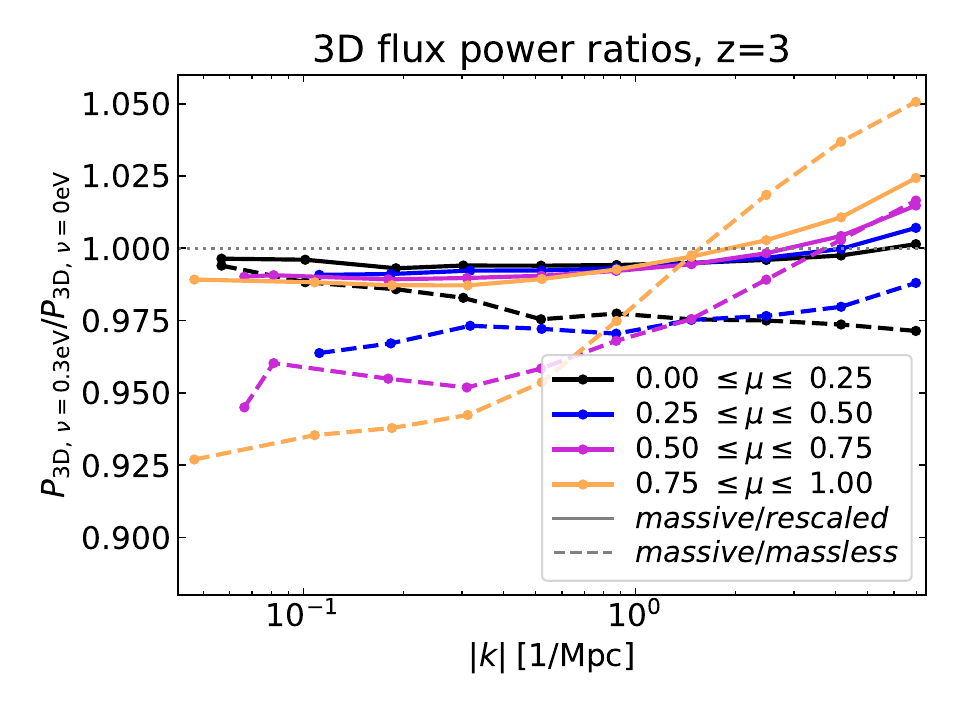}
  \includegraphics[scale=0.47]{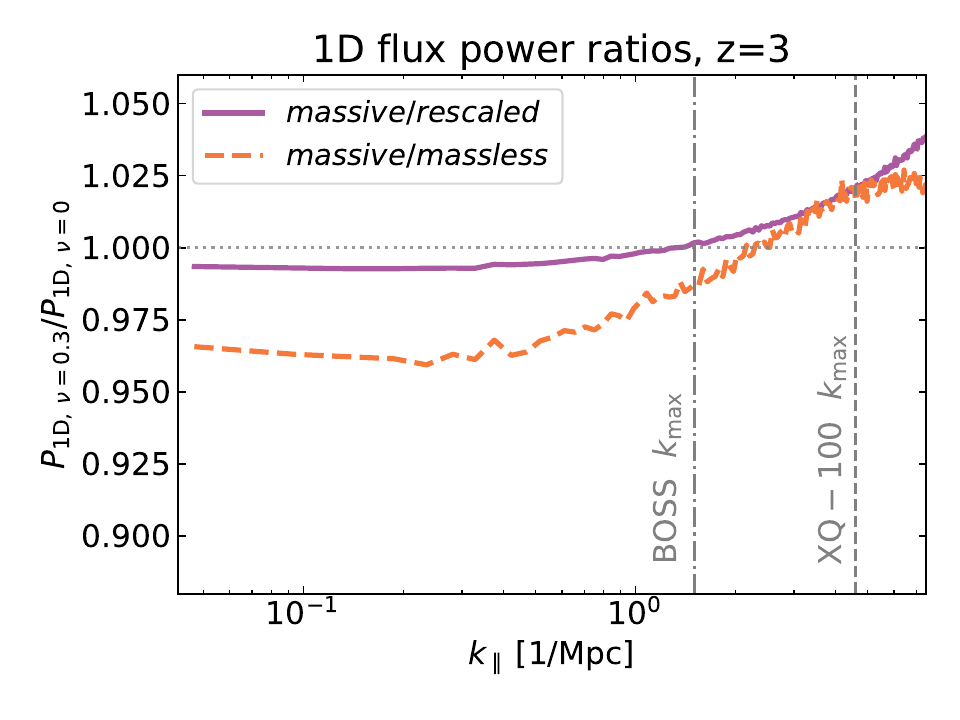}
  \caption{\textit{Left:} Ratios of the 3D flux power spectra in simulations
  with and without massive neutrinos as a function of 3D
    Fourier mode orientation $\mu$.
    Dashed lines show the ratios for the {\it massless} simulation, while
    solid lines show the ratios for the {\it rescaled} simulation. In both
    cases, $T_0$ and the mean flux have been matched in each simulation pair.
  \textit{Right:} Ratios of 1D flux power spectra in massive and
    massless neutrino cosmologies.
    The ratio for the {\it massless} simulation is shown in orange dashed lines
    and the ratio for the {\it rescaled} simulation in purple solid lines.
    In the gray dot-dashed line we show the highest $k_\parallel$ bin measured in
    BOSS data \cite{PD2013}, and in gray dashed we show the
    we show $k_{\mathrm{max}}$ for higher resolution data \cite{Irsic2016}.
    \label{fig:lyaf}
}
\end{figure}

In figure \ref{fig:lyaf} we look at the effect of massive neutrinos on the
flux power spectra after recalibrating $\bar{F}$ and $T_0$ as described above,
and study the degeneracy with the amplitude of the linear power.
The left panel shows the ratios of the 3D flux power spectra in simulations with
and without massive neutrinos. The dashed lines show the comparison between the
{\it massive} and {\it massless} simulations. There is a significant difference
between the flux power spectra in these two simulations, particularly along the line
of sight, with $\mu \sim 1$. The solid lines show that the majority of this difference
comes from the difference in the amplitude of the linear power, as the difference
in the flux power spectra shrinks to $\lesssim 1\%$ for $k<4 \iMpc$ once the amplitude
of the linear power is matched.
In particular for modes transverse to the line of sight (shown in black) there is a
near perfect match. The high $k$ modes along the line of sight still show some deviation,
which is suggesting some residual difference in the thermal state of the IGM.
Additionally there is a
a signature of neutrinos that is not captured in the matter power spectrum
but does affect the flux power spectrum, which comes from the change in the growth
rate. As discussed in section \ref{ss:non}, the presence of massive neutrinos
causes a $2\%$ difference in the growth rate, which has an effect on the flux
power spectrum through a change in the gas velocities. This is a feature of
massive neutrinos that we do not account for in our current setup, but the results
in figure \ref{fig:lyaf} indicate that the effect is very small.

The right panel shows the ratios of the 1D flux power spectra for the same
simulations, where the orange dashed line is the ratio of the {\it massless}
and {\it massive} simulations, and the purple solid line is the ratio of the
{\it massive} and {\it rescaled}. The vertical gray dashed lines show the
highest $k_\parallel$ modes that have been used in recent cosmological
analysis. The purple solid line shows that the effects of massive neutrinos
can be replicated by rescaling the linear power to within at least $1\%$ on scales measured by BOSS,
and within $2.5\%$ for the higher resolution data\footnote{Note that the 10km/s skewer
resolution is low compared with high resolution spectra e.g. \cite{Irsic2016}.
We therefore verified that our results are unchanged when adopting 5 km/s skewers.}.

Upcoming measurements from DESI are not expected
to increase to significantly higher values of $k_\parallel$, but the improvement in
the data will come from a more precise measurement of the same $k_\parallel$ range. 
Therefore the results that we present in this paper will still be applicable
for analysing DESI datasets.
We reiterate that our approach of matching the mean flux and $T_0$ does not 
fully explore the degeneracy. For example, the optimum match of the ratios
shown in figure \ref{fig:lyaf} might be for IGM values that are not the same in each
simulation, i.e. $\bar{F}$ and $T_0$ could be tweaked to push the ratios
closer to 1. What will ultimately matter is to what extent the effects of $\Sigma m_\nu$
are degenerate with a change in these nuisance parameters, which would require
a full marginalisation.

While the results in figure \ref{fig:lyaf} are shown only at $z=3$, in appendix
\ref{app:red} we show the equivalent plots for other redshfits 
and note that the results are largely independent of redshift. Together these results 
indicate a strong degeneracy between $A_s$ and $\Sigma m_\nu$
when considering only the length scales and redshift ranges that are observed using the
\lyaf. 

\section{Conclusion}
\label{sec:con}

We have considered the effects of massive neutrinos on the growth of structure within the length
scales and redshift range relevant for \lyaf\ analysis. These effects can be split
into three categories. First there is a suppression of the overall amplitude
of the power spectrum. Second there is an increase in the non-linear growth caused by the
presence and clustering of massive neutrinos. Third there is an effect
on the growth rate and velocity power spectrum.
The first effect is large, with a $9\%$ suppression of the power spectrum
for $\Sigma m_\nu=0.3$ eV. The \lyaf\ is sensitive to the amplitude of the late-time, small
scale power spectrum, and so this is the strongest signal of neutrino
mass when using \lyaf\ data. However we have shown that this signal is
extremely degenerate with a change in the primordial perturbation amplitude,
$A_s$. In section \ref{ss:non} we showed that even non-linear effects caused by
massive neutrinos are also highly degenerate with a change in the overall
amplitude of the power spectrum.

In figure \ref{fig:relabel} we showed that
the effect of massive neutrinos on the growth rate is small, with a
$<2\%$ effect on the non-linear modes. The \lyaf\ is far less sensitive
to the growth rate - Ref \cite{McDonald2005} measured the growth rate
to a precision of $30\%$, while the amplitude of the linear power was
measured to a precision of $13\%$ - and so we argue that effects on the
growth rate do not break the degeneracy with $A_s$. We demonstrated
this degeneracy on the flux power spectra of the \lyaf\, and showed that,
after matching the mean flux and temperature at mean density,
the effect of $\Sigma m_\nu=0.3$ eV massive neutrinos
is degenerate with $A_s$ to within $<1\%$ in the 1D
flux power at $k_\parallel<3 \iMpc$ and in the 3D flux power at $|k|<4\iMpc$.

Therefore, from the point of view of a \lyaf\ only likelihood,
we have shown that it is not necessary to include an extra parameter to
describe neutrino mass, and that doing so introduces a strong degeneracy.
An interesting future work would be to investigate to what extent
this degeneracy persists into higher order statistics.
Our results also suggest that other parametrisations based on the amplitude
of the linear power at low $z$ and high $k$ would more closely describe the
observables, but we leave the exact specification of such parametrisations
for future work.
We finish by stressing that, even in the presence of this
degeneracy, the \lyaf\ is still a very competitive probe of massive neutrinos
when combined with results from CMB experiments.
The CMB temperature fluctuations depend on the linear power spectrum at
early times, before neutrinos become non-relativistic, and provide a
measurement of $A_s$ that can be used to break the $A_s - \Sigma m_\nu$
degeneracy discussed in this work.
We expect that combined CMB + \lyaf\ analyses will continue to play an
important role in cosmological studies of neutrino masses in the coming years.

\acknowledgments
The authors thank the organisers and participants of the IGM2018 conference at
Kavli IPMU for hosting valuable discussions.
We thank Yu Feng for assistance in running \texttt{MP-Gadget}.
This work was partially enabled by funding from the UCL Cosmoparticle
Initiative.
This work was supported by collaborative visits funded by the Cosmology and
Astroparticle Student and Postdoc Exchange Network (CASPEN).
AFR acknowledges support by an STFC Ernest Rutherford Fellowship, grant
reference ST/N003853/1. 
AFR and AP were further supported by STFC Consolidated Grant number ST/R000476/1.
TDK and AP are supported by a Royal Society University Research Fellowship.
SB was supported by NSF grant AST-1817256.
KKR was supported by the Science Research Council (VR) of Sweden.
This work was performed using the Cambridge Service for Data Driven Discovery
(CSD3), part of which is operated by the University of Cambridge Research
Computing on behalf of the STFC DiRAC HPC Facility (\url{www.dirac.ac.uk}).
The DiRAC component of CSD3 was funded by BEIS capital funding via STFC capital
grants ST/P002307/1 and ST/R002452/1 and STFC operations grant ST/R00689X/1.
DiRAC is part of the National e-Infrastructure.

\appendix

\section{Results at other redshifts}
\label{app:red}

\begin{figure}[h!]
  \centering
  \includegraphics[scale=0.65]{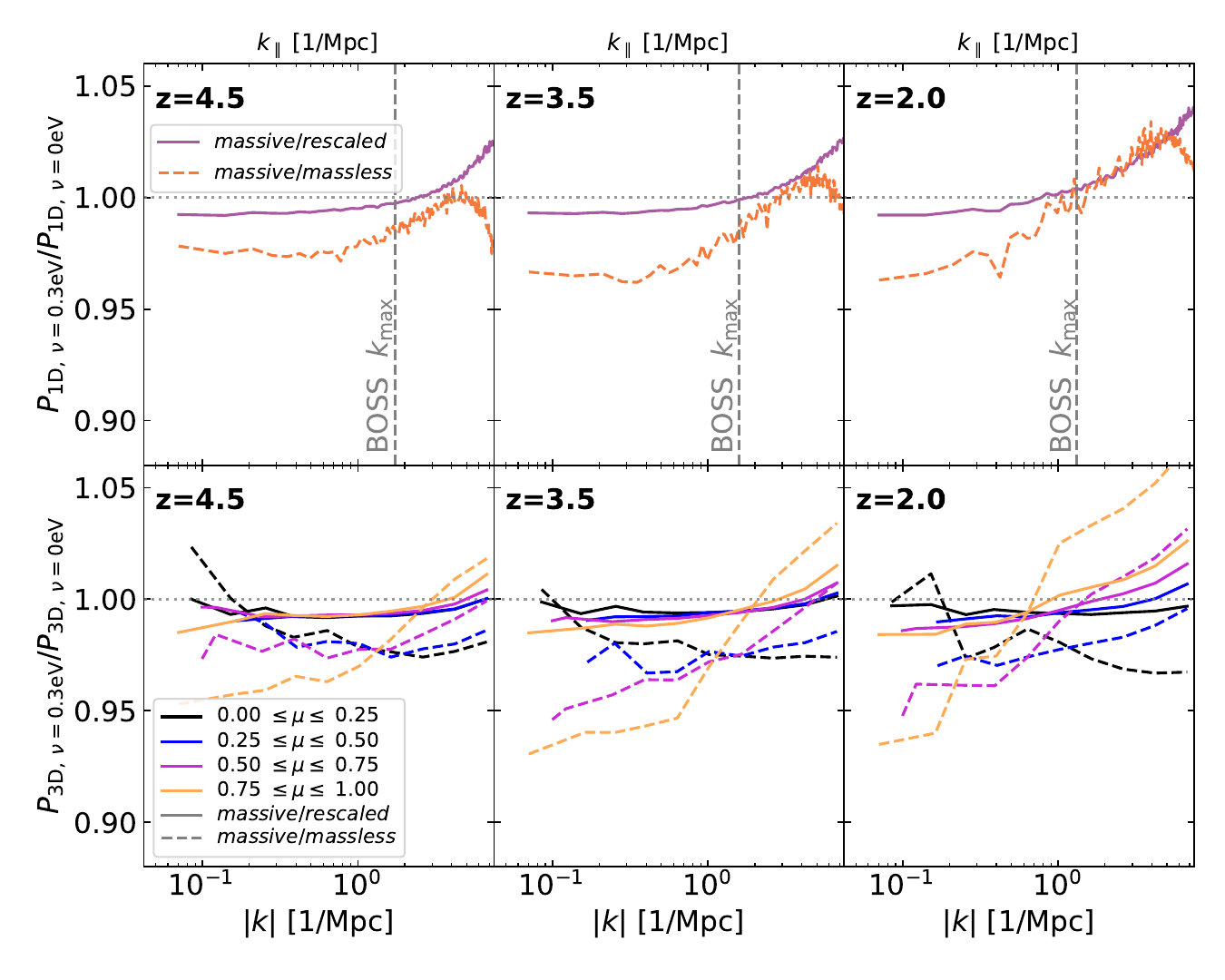}
  \caption{Ratios of 1D (top) and 3D (bottom) flux power spectra in
    simulations with massive and massless neutrinos, at different redshifts.
    In both panels, the solid lines show the ratios of the flux power spectra in the
    {\it massive} simulation and the {\it rescaled}, the ratios for the
    {\it massive} and {\it massless} simulation are shown in dashed lines.
    The mean flux and the temperature at mean density have been matched in
    both simulations to reduce the effect of subtle changes in the thermal
    and ionization history of the IGM. In vertical lines we show the highest $k$
    mode measured from BOSS data in \cite{PD2013} for each redshift bin.
}
  \label{fig:otherz}
\end{figure}

In figure \ref{fig:otherz} we show the ratios of both the 1D and 3D
flux power spectra for the {\it massive} and {\it rescaled} simulations
at redshifts $z=4.5, 3.5, 2$.
We have matched the mean flux
and temperature at mean density. The ratios of the 1D and 3D flux power
spectra have essentially the same amplitude and shape as the $z=3$ ratios presented
in section \ref{ss:lya}. This shows that over the full redshift range relevant
for \lyaf\, the flux power spectrum is primarily dependent on the amplitude
of the linear power and the state of the IGM, and the effects of massive
neutrinos can be replicated to within $2\%$ by matching these parameters.

\section{Dependence on box size and resolution}
\label{app:box}
\begin{figure}[h!]
  \centering
  \includegraphics[scale=0.6]{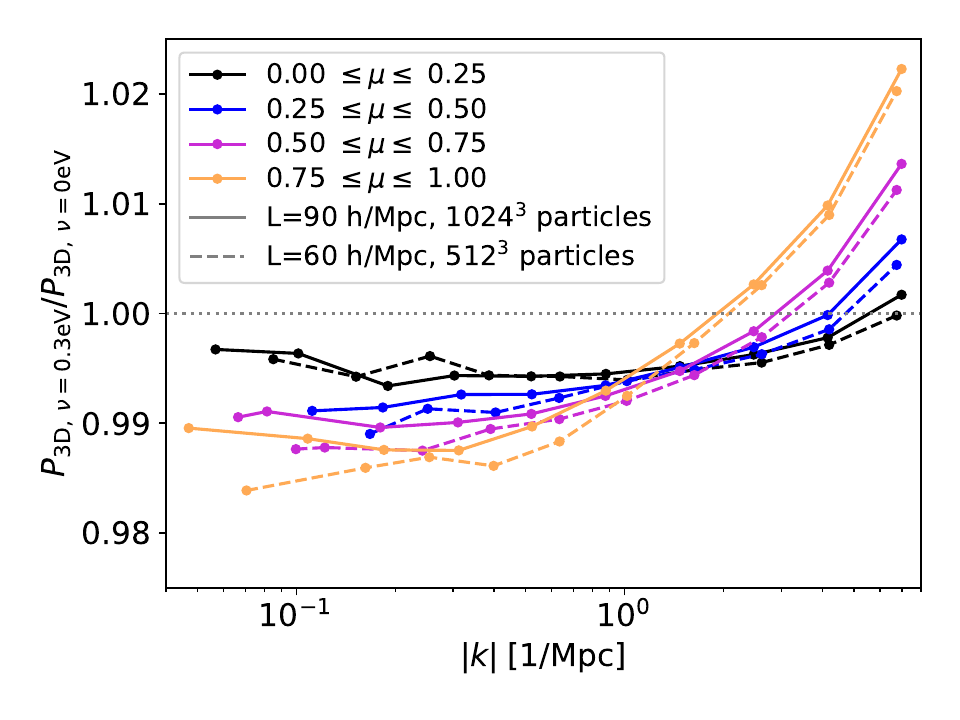}
  \caption{Ratios of the 3D flux power spectra in the {\it massive} and {\it rescaled}
    cosmologies at $z=3$ for two different sets of simulations;
    one with box sizes of $L=90 \hMpc$ and $1024^3$ particles (solid lines),
    and one with box sizes of $L=60 \hMpc$ and $512^3$ particles (dashed lines).
    We have matched $\bar{F}$ and $T_0$ in each simulation pair.
    The results presented in section \ref{ss:lya} are independent of box size
    and resolution.
}
    \label{fig:box}
\end{figure}

In section \ref{sec:res} we have presented the results from a set of
simulations with a box size of $L=133.85 \Mpc$ ($h\,L= 90 \Mpc$), and
$1024^3$ CDM and baryon particles.
In order to test that the main results do not depend on the chosen box size
or resolution, in figure \ref{fig:box} we reproduce the results of the left panel
in figure \ref{fig:lyaf} from simulations with $L=89.23 \Mpc$ ($h\,
L= 60 \Mpc$), and $512^3$ CDM and baryon particles. The original results from the main
text are shown in solid lines, and the dashed lines show the results from the
simulation with lower box size and lower resolution.
There is very little difference between the results shown in the solid and dashed lines,
indicating that the results shown in section \ref{ss:lya} are independent of simulation
box size and resolution.

\section{Dependence on neutrino implementation}
\label{app:nu}

\begin{figure}[h!]
  \centering
  \includegraphics[scale=0.6]{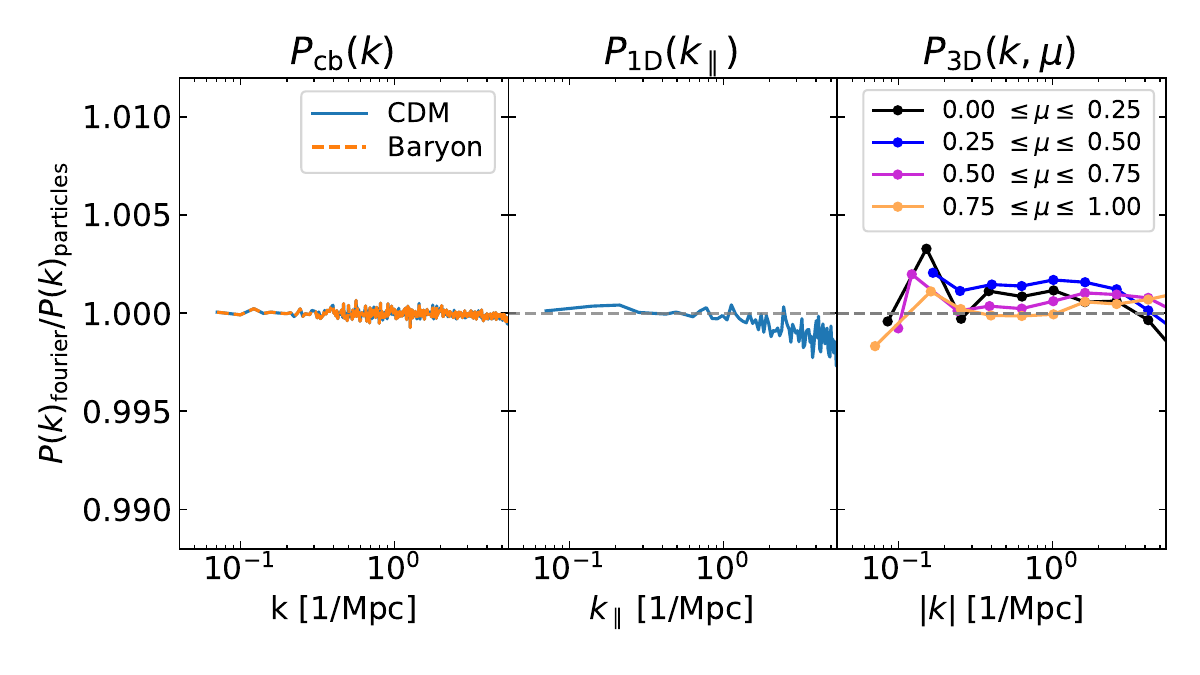}
  \caption{Comparison of the Fourier space based neutrinos with the particle
    implementation at $z=2$.
    From left to right, we show the ratios of the CDM + baryon matter power
    spectra, the 1D flux power and then the 3D flux power spectra.
}
  \label{fig:nu_implementation}
\end{figure}

In order to verify that our results are independent of neutrino implementation,
we show in figure \ref{fig:nu_implementation} the ratio of power spectra at $z=2$
in simulations with the linear response approximation and particle neutrinos.
We include the same number of neutrino particles as CDM and baryons
($N_\mathrm{part}=512^3$), with initial conditions also starting at $z=99$.
The particle implementation can include (small) non-linear clustering
in the neutrino component, whereas the Fourier
space approach only includes non-linearities in the cold dark matter.
We show the comparison at the lowest redshift we have run simulations to as
this is where any disagreement between the two methods would be strongest.
For the CDM + baryon density (left), 1D (center) and 3D (right) flux power
spectra, the agreement between the two approaches is better than $0.5\%$. This is
consistent with the expectation that there is minimal non-linear clustering of
neutrinos at $z=2$ (see refs \cite{Bird2012,Haimoud2013}), and the
residual difference is likely due to shot noise in the neutrino particles.
A similar comparison at other redshifts can be found in figures 5
and 7 of ref \cite{Haimoud2013}.

\bibliographystyle{JHEP.bst}
\bibliography{refs}
\end{document}